# Streaming Potential in Bio-mimetic Microvessels Mediated by Capillary Glycocalyx

Rahul Roy[1], Siddhartha Mukherjee,[2], Rajaram Lakkaraju[1], Suman Chakraborty[1,2*]

[1] *Department of Mechanical Engineering, Indian Institute of Technology Kharagpur, Kharagpur, India-721302*

[2]*Advanced Technology Development Center, Indian Institute of Technology Kharagpur, Kharagpur, India-721302*

Implantable medical devices and biosensors are pivotal in revolutionizing the field of medical technology by opening new dimensions in the field of disease detection and cure. These devices need to harness a biocompatible and physiologically sustainable safe power source instead of relying on external stimuli, overcoming the constraints on their applicability in-vivo. Here, by appealing to the interplay of electromechanics and hydrodynamics in physiologically relevant microvessels, we bring out the role of charged Endothelial Glycocalyx layer (EGL) towards establishing a streaming potential across physiological fluidic conduits. We account for the complex rheology of blood-mimicking fluid by appealing to Newtonian fluid model representing the blood plasma and a viscoelastic fluid model representing the whole blood. We model the EGL as a poroelastic layer with volumetric charge distribution. Our results reveal that for physiologically relevant microflows, the streaming potential induced is typically of the order of 0.1 V/mm, which may turn out to be substantial towards energizing biosensors and implantable medical devices whose power requirements are typically in the range of micro to milli Watt. We also bring out the specific implications of the relevant physiological parameters towards establishment of the streaming potential, with a vision of augmenting the same within plausible functional limits. We further unveil that the dependence of streaming potential on EGL thickness might be one of the key aspects in unlocking the mystery behind the angiogenesis pattern. Our results may open up novel bio-sensing and actuating possibilities in medical diagnostics as well as may provide a possible alternative regarding the development of physiologically safe and biocompatible power sources within the human body.

---

*Corresponding author, email: suman@mech.iitkgp.ernet.in

# I. INTRODUCTION

Recent advancements in the field of micro and nano fluidics have opened up many inroads in the field of medical science and technology. Understanding of blood flow dynamics and its various electrochemical interactions in microvascular environments has been pivotal in addressing several issues related to medical diagnostics and cure (Arun *et al.* 2014; Bandopadhyay *et al.* 2016; Das *et al.* 2006; Das & Chakraborty 2007; Mandal *et al.* 2012). The walls of microvessels are lined by a negatively charged deformable layer of brush like structures called endothelial glycocalyx layer (EGL). This layer consists of many proteins, glycolipids, glycoprotiens and proteoglycans (Pries *et al.* 2000). Although EGL primarily behaves as a mechanotransducer of mechanical stresses to the underlying endothelial cells (Tarbell & Pahakis 2006), its negative charge is responsible for many electrokinetic phenomena arising from its electrochemical interactions with ions dissolved in the blood plasma.

In the microvasculature of human bodies, fluid flow is triggered by pressure-gradient driven pumping action. This may lead to a preferential migration of the ionic species present in the EGL along the downstream direction, leading to the establishment of a streaming potential field under dynamic conditions (Das & Chakraborty 2010; Nguyen *et al.* 2013; Van Der Heyden *et al.* 2005); (Poddar *et al.* 2016). However, the situation gets elusively involved and convoluted, as a consequence of a complicated interplay of the hydrodynamics in a deformable layer and electromechanics amidst complex rheology. While this may turn out to be of significant consequence towards empowering implanted medical devices (Chakraborty 2019), reported literature has turned out to be grossly inadequate towards delineating the underlying implications. This deficit stems from the complexities in coupling the electromechanics and hydrodynamics in biomimetic fluidic pathways. Further, although a few sophisticated experimental studies of flow dynamics near EGL have been carried out (Pries *et al.* 2000; Weinbaum *et al.* 2007), the size and delicate nature of the EGL makes it very difficult to carry out in-vivo experiments, in an effort to probe the underlying electrokinetics.

In the literature, various mechanistic models of the EGL have been reported (Hariprasad & Secomb 2012). The deformability of EGL has been taken into account by considering a biphasic mixture theory (BMT) in several reported studies, where the EGL is modeled as a hydrated porous material with a linearly elastic solid phase and a fluid phase that are spatially coincidental (Damiano & Stace 2005; Lee *et al.* 2016; Sumets *et al.* 2015). These studies have mainly underlined the importance of EGL as a mechanotransducer which transmits most of the mechanical stress through the solid matrix rather than the fluid phase and thus protecting the endothelial surface from excessive fluid shear stress. Although the BMT perfectly outlines the stress distribution aspect of EGL, it considers EGL to be electrically neutral, which represents a situation far from the physical reality. In reality, blood plasma is predominantly a solution of water and various salts containing ions mainly Na+ and $Cl^-$ , whose concentrations are of the same order as that of the fixed charge concentration of EGL (~0.1M) (Silberberg 1991).

Overcoming the above limits, subsequent efforts have endeavored taking the charged nature of the EGL into account. Primarily, three models of EGL have been used to delineate the effects of charged EGL, namely, (a) charged surface model (Liu & Yang 2009), (b) volume charge model (Damiano & Stace 2002; Stace & Damiano 2001), and (c) charged rod model (Buschmann 1995; Mokady *et al.* 1999). EGL structure and mechanics are not considered in the charged surface model, since it assumes zero thickness of the EGL. The charged rod model considers EGL as an array of rigid cylindrical charged rods that are parallel to each other, surrounded by a Newtonian ionic fluid. This model has been used to study electrostatic force characteristics and its effect on streaming potential and electrophoretic mobility of red blood cells (Mokady *et al.* 1999). Though many important conclusions were drawn using charged rod model, the deformability of EGL, which is essential for its role as a mechanotransducer, is not considered in this model.

A study by Damiano & Stace (2002), addressing the motion of a stiff leukocyte over the EGL, concluded that the glycocalyx restoration force is dependent on the ion concentration and thus can be increased or decreased by changing the same. Donath & Voigt (1986), in a different context, concluded that surface conductivity is a direct function of the surface layer thickness, with thick layers showing extraordinarily high surface conductivity. Although these studies have drawn significant conclusions, their coupling with the incipient microvascular flow remained elusive.

Sumets *et al.* (2018), in a pioneering study, extended the work of Damiano & Stace (2002) with the consideration of pressure driven flow of an ionized Newtonian fluid over the poroelastic EGL, and showed the flow reversal near the wall due to ion-EGL electrical interaction. The effects of physiologically relevant thin Debye layer on interfacial interaction between the lumen and EGL were analyzed asymptotically. An important conclusion was drawn that the near-wall flow reversal has its effects on the stress distribution, which eventually affects the EGL's role as a transmitter of mechanical signals from blood flow to the vessel surface. However, this study did not aim to offer any insight on the interplay of electromechanics at deformable interfaces and the rheology of a complex fluid, so as to bring out novel fluid dynamic implications suggesting the possibilities of harnessing electrical energy from the transport of a bodily fluid in microvascular fluidic passages.

Blood is a complex suspension consisting of many particles like red blood cells (RBCs), white blood cells (WBCs), platelets etc., with RBC being the most prominent factor from fluid dynamic purview. Each cellular entity has different shape (RBC biconcave, WBC nearly spherical), size (platelets being the smallest and RBC and WBC being comparatively larger) and deformability (with RBC being deformable and WBC being comparatively rigid) (AlMomani *et al.* 2008; Fung 1993). These different properties result in non-Newtonian behavior of the whole blood. One approach of modeling blood flow is to explicitly resolve its cellular matters (Ding & Aidun 2006; Krüger 2012). However, since these models are computationally expensive, most of the reported studies on blood flow have effectively chosen some representative constitutive models, such as the Power law model (Bandopadhyay & Chakraborty 2011); (Chakraborty 2005; Das & Chakraborty 2006), Carreau (Boyd *et al.* 2007) and Casson (Sankar & Lee 2010) models for inelastic fluids, and Maxwell, UCM, PTT models for viscoelastic fluids, respectively (Afonso *et al.* 2009; (Bandopadhyay & Chakraborty 2012); Bautista *et al.* 2013; Ghosh & Chakraborty 2015; Mukherjee *et al.* 2019; P. Thien 1978; Thien & Tanner 1977), in an effort to bring out the underlying hydrodynamic features. Casson and PTT models have been found to adequately capture the rheological characteristics of blood, and are thus well suited for blood flow problems.

Apart from the non-Newtonian behavior, the deformability characteristics of particles and fluid-solid interaction between the particles and fluid result in phenomena like Fåhraeus and Fåhraeus-Lindqvist effects (Katanov *et al.* 2015; Secomb & Pries 2013; Yin *et al.* 2013). Formation of a cell free layer results in reduced resistance of blood flow through vessels and is also responsible for its shear thinning behavior (Secomb 2016). To take these effects into account, a two fluid model of blood has been assumed in previous studies where blood flow inside the lumen is divided into two regions (Sankar & Lee 2008, 2010). Fluid in the region close to the vessel wall is assumed Newtonian, which is a good approximation of the cell free layer that mainly consists of blood plasma. The fluid in the second region is the whole blood which is modeled by using one of the afore-mentioned non-Newtonian fluid models. These studies have shown that the consideration of the two-fluid model is much more physiologically relevant than a single fluid model. The fluid-fluid interfacial conditions, which are dependent on the physiological properties like hematocrit fraction, have been found to be a significant aspect in determining the blood flow dynamics. Considering that the electrochemical behavior is significantly influenced by the fluid flow dynamics, which is in turn dependent on the complex rheology and physiological parameters, it becomes imperative to take complex rheology into consideration while investigating the potential applications of the electrochemical interaction between EGL and blood.

Here, we bring out the interplay of hydrodynamics and electromechanics in microvessel, in an effort to unravel the corresponding implications on the establishment of streaming potential amidst complex rheological characterizations mimicking a bodily fluid. Towards that, we employ the triphasic mixture theory (TMT), where the physio-chemical theory for the ionic and polyionic mixture and the biphasic mixture theory for porous medium are combined. This results in a poro-elastic EGL model consisting of a fully saturated negatively charged solid skeleton and an incompressible pore fluid containing ions. We consider a two-fluid model for blood (taking whole blood as viscoelastic fluid) for investigating the effects of blood rheology and other physiologically significant parameters on the streaming potential established across the fluidic conduit. By appealing to the constraints of overall electrical neutrality, we characterize the streaming current that is necessary to neutralize the conduction current under a complicated non-trivial interplay between rheologically driven hydrodynamics and electromechanics over interfacial scales. We show that by considering physiologically relevant parameters, streaming potentials compatible with the requirements of empowering implantable medical devices and sensors can be realized, bearing significant implications in medical technology.

## II. PROBLEM FORMULATION

The schematic of the problem is depicted in figure 1 where the flow domain is divided into three regions, namely, region I, II and III respectively. For the symmetry of the problem, only half of the entire domain is considered for the analysis. The region 3 is the Endothelial Glycocalyx layer (EGL) which is treated as a negatively charged poroelastic layer. The volume charge model, similar to the earlier studies, is used to model the poroelastic layer. The thickness of the EGL is taken as ~ 20% of the vessel radius for our analysis.

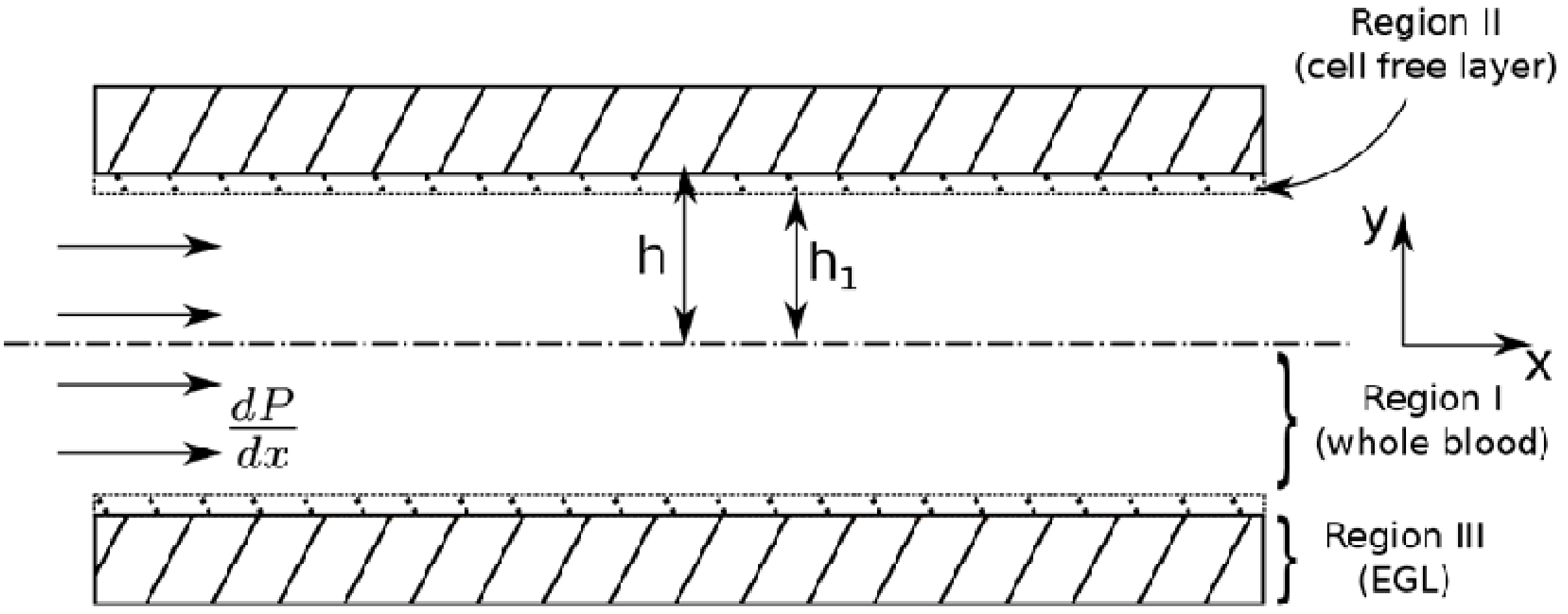

Figure 1: Schematic representation for the blood flow over Endothelial glycocalyx layer (EGL).

Because of the complex rheological behavior of blood, a two-fluid model is assumed where the region II, as shown in figure 1, consists of Newtonian fluid mimicking blood plasma. Region I consists of whole blood which is modeled as viscoelastic fluid. The EGL region is assumed to be consisting of two phases, solid and fluid phase, with volume fractions $\phi_s$ and $\phi_f$ respectively. Water and ions constitute the fluid phase and electrostatically charged elastic solid matrix constitutes the solid phase. By mass conservation, it is implied that $\phi_s + \phi_f = 1$. Because of the physiologically low value of salt concentration in blood plasma, the ions are treated as point charges, resulting in fluid volume fraction almost equal to the water volume fraction. The electrostatic charge of the EGL is characterized by a constant volumetric charge concentration, $c_s$. This volumetric distribution of the charge concentration implies zero surface charge concentration on the interface between the lumen and the EGL.

The flow is assumed to be pressure driven (with a pressure distribution assumed to be of the form $p(x) = Ax$), neglecting the pulsatile nature of real physiological flow as the Womersley number (*Wo*) associated with the flow in capillaries comes out be very less than 2, of the order of ~ $10^{-2}$ (Akbarzadeh 2016). Along with this, the flow is also assumed to be incompressible and fully developed with characteristic velocity of the order of $10^{-3}$ m/s. Vessel radius of $5\times10^{-6}$ m along with the characteristic apparent viscosity of the whole blood results in Reynolds number of the order of $10^{-3}$. Thus, inertial terms in the momentum equation can be neglected safely. The flow is such that the ion concentration at the inlet and outlet are the same, resulting in zero contribution of ion concentration gradient in the flow direction (Sumetc 2017). The properties of Newtonian fluid in region II are taken to be those corresponding to blood plasma at 37 $^{o}$C. Typical values of the properties of whole blood and blood plasma are tabulated below:

| Properties | Values |
|---|---|
| Characteristic velocity | $10^{-3}$ ms$^{-1}$ |
| Vessel radius | $5 \times 10^{-6}$ m |
| Viscosity of whole blood (at 37$^{0}$C) | 16.9 mPa.s |
| Viscosity of blood plasma (at 37$^{0}$C) | 1.34 mPa.s |
| Surface tension of blood (at 37$^{0}$C) | $52 \times10^{-3}$ N/m |

Table 1: Properties of whole blood and blood plasma.

Due to the low ion concentration in plasma, the EDL formed is very thin and of the order of 1 nm (Sumets *et al.* 2018). The cell free layer thickness, i.e. thickness of region II, is physiologically 20-40% of the vessel radius which comes out to be of the order of 1-2 μm (Katanov *et al.* 2015; Secomb 2016). This implies that the EDL is very thin in comparison with the cell free layer thickness thus we can assume that there is no electrical interference between the fluids in region II and I across their interface. The viscosity of whole blood is taken as 16.9 mPa.s (Brust *et al.* 2013) and the characteristic velocity $(V)$ of the flow is $10^{-3}$ m/s (Sumets *et al.* 2018). With the surface tension $(\sigma)$ of blood taken as $52\times10^{-3}$ N/m (Hrnčíř & Rosina 1997), the capillary number $\left( Ca = \frac{\mu_2 V}{\sigma} \right)$ comes out to be of the order of $10^{-4}$. This value is low enough to assume that there is no deformation of the interface between the two fluids and it effectively remains flat (Mandal *et al.* 2015).

With all these assumptions, we will briefly express the equations governing the flow in each region in both dimensional and non-dimensional forms. Subsequently, the boundary and interfacial conditions required to solve the flow equations to obtain the streaming potential are mentioned.

**Region I (Whole blood)**

To describe the rheological behavior of whole blood, we have chosen the constitutive form of the simplified Phan-Thien Tanner model (sPTT), typically used for viscoelastic fluids. In sPTT model, the stress tensor $\boldsymbol{\tau}$ takes the following form (Afonso *et al.* 2009; Bautista *et al.* 2013 ) (Mukherjee *et al.* 2017b)

$$f\left(\tau_{kk}\right)\boldsymbol{\tau}+\lambda_R \overset{\nabla}{\boldsymbol{\tau}}=2\mu_1 \boldsymbol{D} \tag{1}$$

where $f\left(\tau_{kk}\right)$ is the stress coefficient function, $\lambda_R$ is the relaxation time of the fluid, $\mu_1$ is the viscosity of whole blood, $\boldsymbol{D}=\frac{1}{2}\left(\nabla\boldsymbol{u}+\nabla\boldsymbol{u}^T\right)$ is the deformation tensor, $\overset{\nabla}{\boldsymbol{\tau}}=\frac{D\boldsymbol{\tau}}{Dt}-\nabla\boldsymbol{u}^T\cdot\boldsymbol{\tau}-\boldsymbol{\tau}\cdot\nabla\boldsymbol{u}$ is the upper convective derivative, $\tau_{kk}$ is the trace of stress tensor. Also, we have chosen the linear approximation of $f\left(\tau_{kk}\right)$, i.e. $f\left(\tau_{kk}\right)=1+\frac{\in\lambda_R}{\mu_1}\tau_{kk}$ where $\in$ is the extensibility of the fluid. For steady, laminar, incompressible and fully developed flow assumption, the stress tensors are simplified and one can establish a relationship between the tangential stress $\left(\tau_{xy}\right)$ and normal stress $\left(\tau_{xx}\right)$ as $\tau_{xx1}=\frac{2\lambda_R}{\mu_1}\tau_{xy1}{}^2$ (this derivation is shown in **Section A** of the **Appendix**) (Bautista *et al.* 2013; Ferrás *et al.* 2016; Mukherjee *et al.* 2017a). Considering this, the simplified momentum equations in region 1 are given as follows:

$$\left.\begin{aligned} x-\text{component}:\ 0&=-\frac{\partial p}{\partial x}+\frac{\partial\tau_{yx1}}{\partial y}\\ y-\text{component}:\ 0&=-\frac{\partial p}{\partial y}\end{aligned}\right\} \tag{2}$$

From this, we obtain the expression of $\tau_{xy}$ as $\tau_{xy1}=\frac{dp}{dx}y+c_1$ where $\tau_{xy}$ is also related to the velocity gradient as

$$\frac{\partial u_1}{\partial y}=\frac{\tau_{xy1}}{\mu_1}+\frac{2\in \lambda_R^2}{\mu_1^3}\tau_{xy1}^{\ \ 3} \tag{3}$$

(the derivation is shown in **Section A**). The vessel radius $(H = 5\ \mu\text{m})$ (Sumets *et al.* 2018) is taken as the length scale and characteristic blood velocity is taken as the velocity scale $(V = 5\times10^{-4}\ \text{ms}^{-1})$. The viscosity of blood plasma (considered as Newtonian fluid) is taken as the reference viscosity $(\mu_2 = 1.34\ \text{mPa.s})$ (Brust *et al.* 2013) and the salt concentration in blood plasma in taken as the reference ion concentration $(c_0 = 0.154\,\text{mol}\,\text{l}^{-1})$ (Sumets *et al.* 2018). The other variables used in the analysis are non-dimensionalized as follows:

$$p=\frac{\mu_2 V}{H}\overline{p},\ \tau_{ii}=\frac{\mu_2 V}{H}\overline{\tau}_{ii},\ \varphi=\frac{k_B T}{e}\overline{\varphi},\ c=c_0\overline{c} \tag{4}$$

where $e$ is the elementary charge, $k_B$ Boltzmann constant and $T$ absolute temperature.

After making Eq. (2) dimensionless and substituting equation (3), we get the following form of the velocity distribution:

$$\overline{u}_1=\left(\frac{A\overline{y}^2}{2}+c_1\overline{y}\right)\mu_r+2\in De^2\mu_r^3\left(\frac{A^3\overline{y}^4}{4}+c_1^3\overline{y}+A^2\overline{y}^3c_1+\frac{3A\overline{y}^2c_1^2}{2}\right)+c_2 \tag{5}$$

where $A=\frac{d\overline{p}}{d\overline{x}}$, $De$ is Deborah number, i.e. $De=\frac{\lambda_R V}{H}$ and $\mu_r$ is viscosity ratio, i.e. $\mu_r=\frac{\mu_2}{\mu_1}$.

The relaxation time $(\lambda_R)$ of whole blood is chosen as ~ 7 ms (Brust *et al.* 2013) such that the value of Deborah number $(De)$ comes out to be ~ 0.7 for our analysis. For this value of $De$, the linear approximation of the sPTT model gives results which are in agreement with that of the exponential sPTT model (Bautista *et al.* 2013; Mukherjee *et al.* 2017b). Thus, we have considered the linear sPTT model.

At the centreline the velocity equals the characteristic blood flow velocity $(V)$. Therefore, at $\overline{y}=0$, $\overline{u}_1=1$ and $\overline{u}_1'=0$. Applying these conditions yields the velocity profile in region I as

$$\overline{u}_1=\frac{1}{2}A\overline{y}^2\mu_r+\frac{1}{2}\in De^2\mu_r^3A^3\overline{y}^4+1 \tag{6}$$

**Region II (Blood plasma)**

For the cell-free layer, the governing equations for momentum, charge distribution and potential distribution are as follows (Sumets *et al.* 2018):

$$\left.\begin{array}{ll}\text{Momentum Equation:} & \mu_f \nabla^2 u_2 = \nabla p_l + \nabla\left(c_l^+ + c_l^-\right)k_B T + e\left(c_l^+ - c_l^-\right)\nabla\varphi_l \\ \text{Nernst-Planck Equation:} & u_2 \nabla c_l^{\pm} - D_{\pm}\nabla^2 c_l^{\pm} \mp \dfrac{eD_{\pm}}{k_B T}\nabla\left(c_{l\pm}\nabla\varphi_l\right) = 0 \\ \text{Gauss Law:} & -\varepsilon\nabla^2\varphi_l = e\left(c_l^+ - c_l^-\right)\end{array}\right\} \quad (7)$$

where $\mu_f$ is the viscosity of blood plasma which is present in region 2, $p_l$ is the hydrodynamic pressure, $c$ denotes ion concentration where superscripts '+', '-' indicate positive and negative ions respectively. $D$ is the diffusivity of the ions, $\varepsilon$ is electrical permittivity and $\varphi$ denotes the electric potential. The subscript '*l*' represents lumen. The non-dimensional electric potential $\bar{\varphi}$ is expressed as the summation of two potentials as $\bar{\varphi}(\bar{x},\bar{y}) = \bar{\psi}(\bar{y}) + B\bar{x}$ where $\bar{\psi}(\bar{y})$ is the potential distribution within the EDL and $B$ is the spatially uniform electric field strength due to streaming potential. Using this and the above mentioned non-dimensionalisation and flow assumptions, the following equations are obtained:

$$\left.\begin{array}{c}\bar{u}_2'' = A + \hat{\chi}_l\left(\bar{c}_l^+ - \bar{c}_l^-\right)B \\ \left[-\bar{c}_{l\pm} \mp \bar{c}_{l\pm}\bar{\psi}_l'\right]' = 0 \\ -\bar{\psi}_l'' = \lambda\left(\bar{c}_l^+ - \bar{c}_l^-\right)\end{array}\right\} \quad (8)$$

where is $\hat{\chi}_l$ the Hartmann number (*Ha*) which represents the Coulomb body force in the lumen and is given by $\hat{\chi}_l = \dfrac{c_0 k_B T H}{\mu_2 V}$ and $\sqrt{\lambda}$ is the inverse of the Debye layer thickness given by $\lambda = \dfrac{c_0 e^2 H^2}{\varepsilon k_B T}$.

**Region III (EGL)**

Considering EGL as a poroelastic layer, as modeled by Sumets *et al.* (2018), the following governing equations are obtained:

$$\left.\begin{aligned}
&\text{Momentum Equations: 1.} \quad \mu_f \phi_f \nabla^2 u_3 = \phi_f \nabla\left\{p + \left(c^+ + c^-\right) k_B T\right\} + \phi_f e\left(c^+ - c^-\right)\nabla\varphi + K u_3 \quad (\text{fluid}) \\
&\qquad\qquad\qquad\qquad \text{2.} \quad \phi_s\left(\lambda_s + \mu_s\right)\nabla^2 u_s = \phi_s \nabla\left\{p + \left(c^+ + c^-\right) k_B T\right\} + \phi_s\, e\, z_s c_s \nabla\varphi - K u_3 \quad (\text{solid}) \\
&\text{Nernst - Planck Equation:} \quad \nabla\cdot\left(u_3\, c_\pm - D_\pm \nabla c_\pm - \frac{e\nabla\varphi}{k_B T} z_\pm D_\pm c_\pm\right) = 0 \\
&\text{Gauss Law:} \qquad -\varepsilon\nabla^2\varphi = e\left(c^+ - c^- + z_s c_s\right)
\end{aligned}\right\} \quad (9)$$

and the simplified dimensionless forms after inculcating the flow assumptions are as follows:

$$\left.\begin{aligned}
&\bar{u}_3'' = A + B\hat{\chi}\left(\bar{c}^+ - \bar{c}^-\right) + \chi\bar{u}_3 \quad (\text{fluid}) \\
&\bar{u}_s'' = \phi A - B\hat{\chi}\bar{c}_s - \chi\bar{u}_3 \qquad (\text{solid}) \\
&\left(-\bar{c}_\pm' \mp \bar{c}_\pm \bar{\psi}'\right)' = 0 \\
&-\bar{\psi}'' = \lambda\left(\bar{c}^+ - \bar{c}^- - \bar{c}_s\right)
\end{aligned}\right\} \quad (10)$$

The subscript '$s$' refers to the properties of the solid matrix and $\chi$ is the Darcy permeability. $\bar{u}_3$ indicates the fluid velocity inside the EGL. Here, $\bar{u}_s$ indicates that the deformation of the EGL does not have any effect on the streaming potential and thus, we have only focused on the fluid velocity in the three regions which influences the value of streaming potential. However, $\bar{u}_s$ can be calculated separately after determining the pressure gradient and streaming potential by using the associated boundary condition. For our convenience, from now onwards we will use the non-dimensional equations without bars.

The boundary conditions to solve the flow equations are as follows:

(i) At the vessel wall, i.e., at $y = \pm 1$, $u_3 = 0,\ \psi' = 0,\ c_\pm' = 0$ (11)

(ii) At the interface between region 3 and region 2, i.e.,

at $y = \pm h = \pm(1-\varepsilon)$, $u_2 = \phi_f\, u_3, u_2' = u_3', \psi_l = \psi, \psi_l' = \psi'$ and $c_{l\pm} = c_\pm$ (12)

By solving the governing equations (8) and (10) of the ion concentration using the boundary conditions as mentioned in equation (11) and equation (12), we obtain $c_{l\pm} = e^{\mp\psi_l}$ and $c_\pm = e^{\mp\psi}$, respectively. By substituting the ion concentration expression in equation (8) and (10), one can determine the potential distribution $\psi$ as

$$\left.\begin{aligned}\psi_l'' &= 2\lambda \sinh(\psi_l) \\ \psi'' &= 2\lambda \sinh(\psi) + \lambda c_s\end{aligned}\right\} \tag{13}$$

In the earlier studies, it has been found that the assumption of small electric potential agrees quite well with that of full numerical solution. Thus, we arrive at the following system of linear equations by assuming small values of electric potential

$$\left.\begin{aligned}\psi_l'' &= 2\lambda \psi_l \\ \psi'' &= 2\lambda \psi + \lambda c_s\end{aligned}\right\} \tag{14}$$

After solving these equations by applying the boundary conditions, we obtain

$$\left.\begin{aligned}\psi_1 &= K \cosh\left(\sqrt{2\lambda} y\right) \\ \psi &= K_1 e^{-\sqrt{2\lambda} y} + K_2 e^{\sqrt{2\lambda} y} - \frac{c_s}{2}\end{aligned}\right\} \tag{15}$$

where $K_1 = K_2 e^{2\sqrt{2\lambda}}$, $K_2 = -\dfrac{K \sinh\left(\sqrt{2\lambda}\, h\right)}{2e^{\sqrt{2\lambda}} \sinh\left\{\sqrt{2\lambda}(1-h)\right\}}$ and $K = -\dfrac{c_s \sinh\left\{\sqrt{2\lambda}(1-h)\right\}}{2\sinh\left(\sqrt{2\lambda}\right)}$.

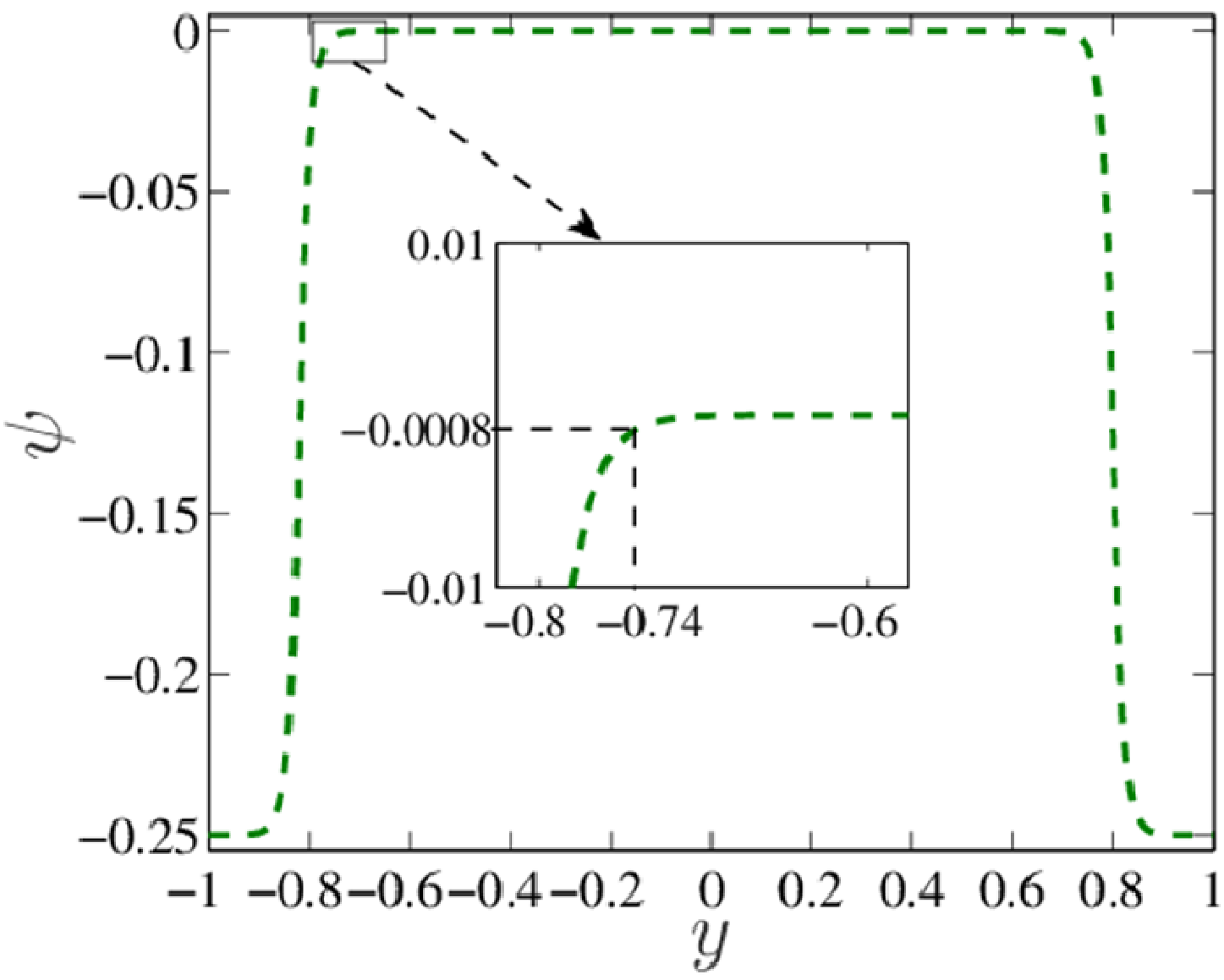

Figure 2: Potential distribution along the transverse direction inside the microvessel. A magnified view of a portion is shown in inset to highlight the fact that the potential value decreases to zero after $y = -0.74$.

We have chosen the value of $\lambda$ as $2\times10^4$ which results in Debye layer thickness of 30 nm. This value of $\delta$ is still very small as compared to the cell-free layer thickness which is 20 to 40% of the vessel radius, i.e., 1 to 2 μm for our case. Thus, the electrical double layer (EDL) will still remain inside the cell free layer and will not interact with the whole blood. Accordingly, it is safe to assume that no electrochemical interaction takes place between the whole blood and blood plasma and hence, Maxwell stress balance need not be invoked at the fluid-fluid interface. The potential distribution for $\lambda = 2\times10^4$ is shown in Fig. 2, which shows that the potential decreases as we move from the wall towards the centreline and becomes zero after $y = 0.74$ (i.e. inside the cell-free layer).

Now, by using the above expressions of potential and ion concentration distribution, the velocity profile in region 2 and region 3 can be expressed as:

$$\left.\begin{aligned} u_2 &= \frac{1}{2}Ay^2 - \frac{\hat{\chi}_l BK}{\lambda}\left[\cosh\left(\sqrt{2\lambda}\,y\right)\right] + c_1 y + c_2 \\ u_3 &= D_2 e^{\sqrt{\chi}\,y} + D_1 e^{-\sqrt{\chi}\,y} + \frac{2B\left(K_1 e^{-\sqrt{2\lambda}y} + K_2 e^{\sqrt{2\lambda}y}\right)}{\chi - 2\lambda} - \frac{A + Bc_s\hat{\chi}}{\chi} \end{aligned}\right\} \tag{16}$$

Now, the velocity profile of region 2 is further solved by using the interfacial boundary conditions between region 2 and region 1, i.e., at $y = h_1$, $u_1 = u_2$ and $\tau_{xy1} = \frac{\partial u_2}{\partial y}$. After applying these boundary conditions, we obtain the complete velocity profile in region 2 in terms of the pressure gradient $(A)$ and streaming potential $(B)$, as

$$\left.\begin{aligned} u_2 = &\frac{1}{2}Ay^2 - \frac{\hat{\chi}_l BK}{\lambda}\left[\cosh\left(\sqrt{2\lambda}\,y\right)\right] + \frac{\hat{\chi}_l BK\sqrt{2\lambda}}{\lambda}\left[\sinh\left(\sqrt{2\lambda}\,h_1\right)\right]y + \frac{Ah_1^2}{2}(\mu_r - 1) \\ &+ 1 + \frac{\in}{2}De^2\mu_r^3 h_1^4 A^3 + \frac{\hat{\chi}_l BK}{\lambda}\left[\cosh\left(\sqrt{2\lambda}\,h_1\right) - h_1\sqrt{2\lambda}\sinh\left(\sqrt{2\lambda}\,h_1\right)\right] \end{aligned}\right\} \tag{17}$$

We have 4 unknowns $A$, $B$, $D_1$ and $D_2$ which can be determined from the velocity distribution by using the following four constraints:

(i) At $y = 1$, $u_3 = 0$ (no-slip at vessel wall),

(ii) At $y = h$, $u_2 = \phi_f u_3$ (interfacial condition between EGL and cell-free layer)

(iii) At $y = h$, $u_2' = u_3'$ (continuity of viscous stress)

(iv) $\int_{-1}^{-h} i_{EGL}\, dy + \int_{-h}^{-h_1} i_l\, dy + \int_{h_1}^{h} i_l\, dy + \int_{(1-\varepsilon)}^{1} i_{EGL}\, dy = 0$ (zero net axial current, or equivalently, the electroneutrality constraint) where $i_{EGL} = u_3\left(c_+ - c_-\right) - B\left(\frac{c_+}{\gamma_+} + \frac{c_-}{\gamma_-}\right)$ and $i_l = u_2\left(c_{l+} - c_{l-}\right) - B\left(\frac{c_{l+}}{\gamma_+} + \frac{c_{l-}}{\gamma_-}\right)$. Here, $\gamma_+$ and $\gamma_-$ represents the ionic Peclet numbers for cation and anion respectively, $\gamma_\pm = \frac{VH}{D_\pm}$.

After applying these boundary conditions, we get the following cubic equation of the pressure distribution which is solved numerically :

$$A\,C_{27} + A^3\,C_{28} + C_{29} = 0 \tag{18}$$

Along with this, we also obtain a relationship between the streaming potential and the pressure gradient, in the following form:

$$B = A^3 C_{18} + A C_{17} + C_{19} \tag{19}$$

The coefficients of these equations can be found in **Section B** of the Appendix.

## III. RESULTS AND DISCUSSIONS

Because of the fixed negative charge concentration associated with the EGL, the electrostatic potential is negative within the EGL. As we move away from the EGL, a sharp transition of potential is observed inside the cell-free layer, i.e. the layer predominantly containing blood plasma. This is quite evident from the potential distribution curve as shown in figure 2. There after the potential becomes zero in most part of the bulk indicating non-existence of electrical effects. The transition from the negative zeta potential to zero potential in the lumen is characterized by the common phenomena of formation of electric double layer (EDL)(Das *et al.* 2012; Das & Chakraborty 2011). The electrical effects due to the interaction between charged EGL and the electrolyte (blood plasma) are mainly found in the EDL and it diminishes as one move away from it.

The counter-ions (positive ions in this case) present in the plasma moves towards the charged EGL forming an electric screen over the charged EGL. This layer of counter-ions is loosely associated to the EGL surface allowing them to move freely under the influence of external body force. Because of the applied pressure gradient, these free counter-ions are dragged along the downstream direction. This results in accumulation of counter-ions downstream causing an imbalance in ion-concentration which leads to an induced electric potential called streaming potential. The counter-acting streaming potential tends to move the counter-ions in the opposite direction (upstream) resulting in overall reduction of the flow velocity. This is quantified by an apparent increase in the viscosity of the fluid, known as electroviscous effect. In figure 3, a comparison between velocity profile with and without presence of charge in EGL is considered, i.e. a comparison between hydrodynamics in the presence and absence of electrical effects is drawn. It is quite evident from the figure that the magnitude of the flow velocity is lower in presence of electrical effects.

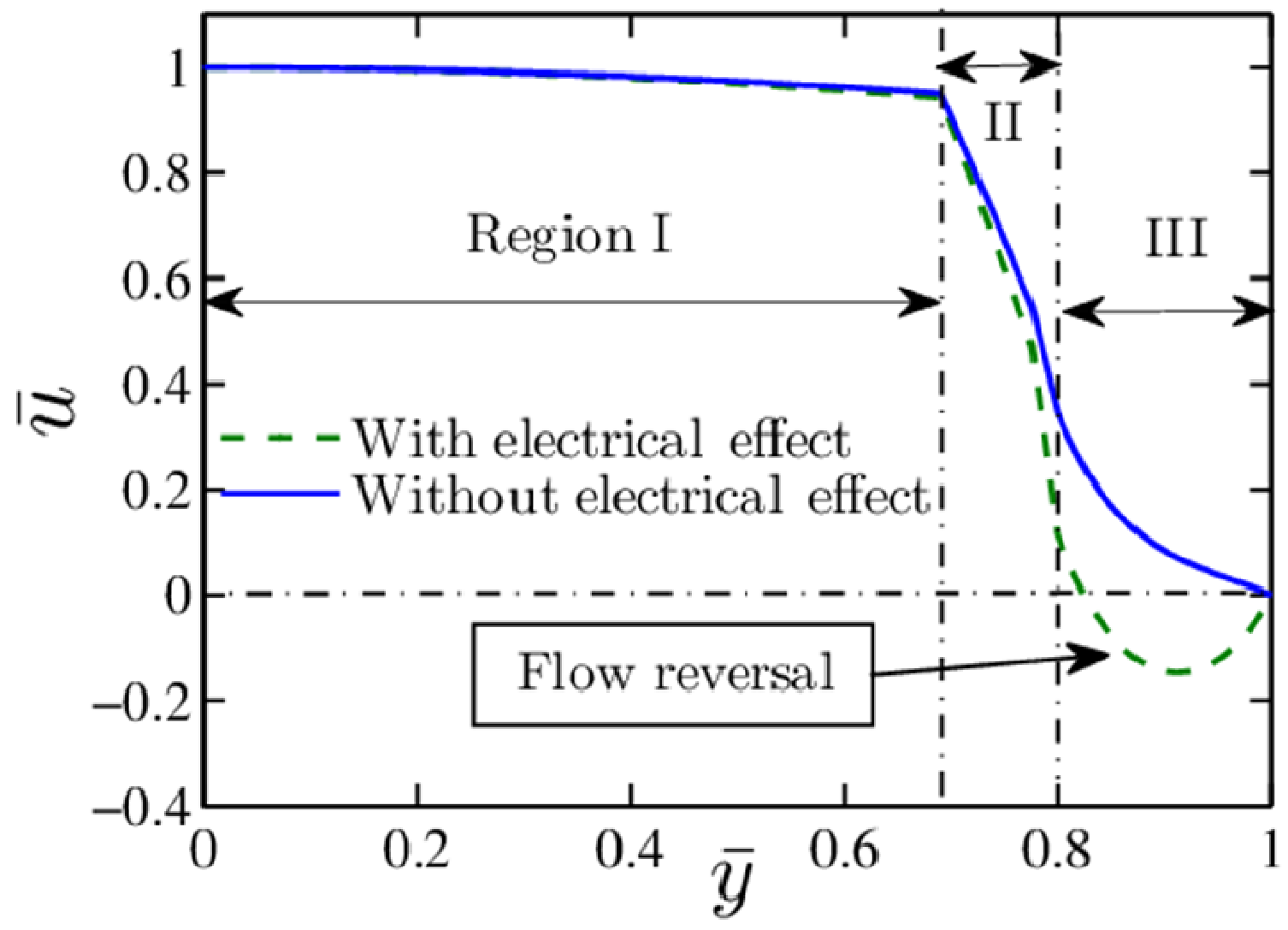

Figure 3: The variation of flow field in the transverse direction both in presence and absence of electric field.

As discussed earlier, electrical effect is mainly dominant within EGL (region III) and gradually diminishes away from the EGL, one can expect the velocity field in region I (whole blood) to be parabolic in nature solely governed by pressure gradient. However, since the viscosities of whole blood (region III) and blood plasma (region II) are considerably different (the ratio is approximately ~ 16:1), in order to maintain the continuity of viscous stress, sharp alteration in the velocity gradient can be observed at the interface between whole blood and blood plasma thereby creating deviation from parabolic nature.

In region II, slight variation in flow field is found in presence of electrical effect which creates a backward flow because of the induced streaming field. So, the net flow depends on the relative contribution between two counter-acting driving forces, which are pressure-gradient and streaming potential induced back flow. Since the strength of streaming field is relatively lower in region II, the effect on flow field is insignificant. However, the induced streaming field has its maximum effect within region III and as a results, a flow reversal is observed within EGL and the EGL cell-free layer interface. This reversal in flow is due to the fact that in this region the back flow associated to the induced streaming potential is higher than the downstream pressure driven flow. The competition between these two aforementioned opposing forces introduces non-linearity in the flow field. Along with this, on closely observing the constitutive behavior of whole blood (i.e. region 1) equation (3), it is evident that hydrodynamic stresses (here both normal and tangential stress exist unlike blood plasma which is Newtonian) are coupled with the velocity gradient thereby inherently introducing non-linearity in the flow field. To maintain the continuity of flow velocity and viscous stress throughout the vessel the non-linear effects induced in the whole blood due to its complex rheology is propagated in region II as well through the interfacial boundary conditions.

While determining the streaming potential by fixing the characteristic centre-line flow velocity, this non-linearity induced in the flow due to combined electroviscous and rheological effects are accounted in the electroneutrality condition i.e. $\int_{-1}^{-h} i_{EGL}\, dy + \int_{-h}^{-h_1} i_l\, dy + \int_{h_1}^{h} i_l\, dy + \int_{(1-\varepsilon)}^{1} i_{EGL}\, dy = 0$ which involves the flow velocity ($i_{EGL} = u_3\left(c_+ - c_-\right) - B\left(c_+/\gamma_+ + c_-/\gamma_-\right)$, $i_l = u_2\left(c_{l+} - c_{l-}\right) - B\left(c_{l+}/\gamma_+ + c_{l-}/\gamma_-\right)$). Thus, to compute the streaming potential in case of blood flow over a charged poroelastic EGL, it becomes necessary to consider a coupled system where electrostatics, hydrodynamics and complex rheology are all interlinked with each other. These effects individually are dependent upon various physiological aspects like hematocrit fraction.

Subsequently, we have first shown the validity of the present analysis by comparing our results with the findings of Summets et al. (2015) which were limited to Newtonian fluid considerations for the entire domain. Then, we have highlighted the importance of two-fluid model over fully Newtonian fluid model as far as proper estimation of streaming potential is concerned. Later, we have discussed the effects of physiologically relevant parameters like hematocrit fraction (Hct), cell free layer thickness and EGL thickness on streaming potential and its possible application in the field of medical diagnostics and cure. It is also shown that the pressure gradient is dependent on these parameters, modulating the induced streaming potential to a significant extent.

Figure 4 depicts a comparison between the values of streaming potential obtained by the consideration of full Newtonian fluid and two fluid model, respectively. This is done to validate the two fluid model with the results obtained by a previously deployed single fluid model (full Newtonian). For this, the Deborah number (*De*) is taken as zero since the relaxation time for a Newtonian fluid is zero. The value of $h_1$ i.e. the distance of fluid-fluid interface from center is taken as zero and the viscosity ratio is taken as 1. Other variables have same values for both the cases. It is found that with these values, the results obtained for the two models are in excellent agreement with each other.

In figure 5, the importance of using the two-fluid model is shown where the viscosity ratio is shown to have a significant influence on the value of streaming potential. The ratio of viscosity between blood plasma and the whole blood at 37 $^{o}$C and 45% hct is found to be close to 0.078 (Brust *et al.* 2013). Taking this value of viscosity ratio, the value of streaming potential is found to be higher than that for a single fluid model (i.e. viscosity ratio = 1). At $c_s = 1$, the streaming potential for the two-fluid model is found to be almost 40% higher than the single fluid model. Thus, it is imperative to consider a physiologically relevant two fluid model for analyzing the effects of various physiological parameters on streaming potential, in an effort to avoid under-prediction of the streaming potential.

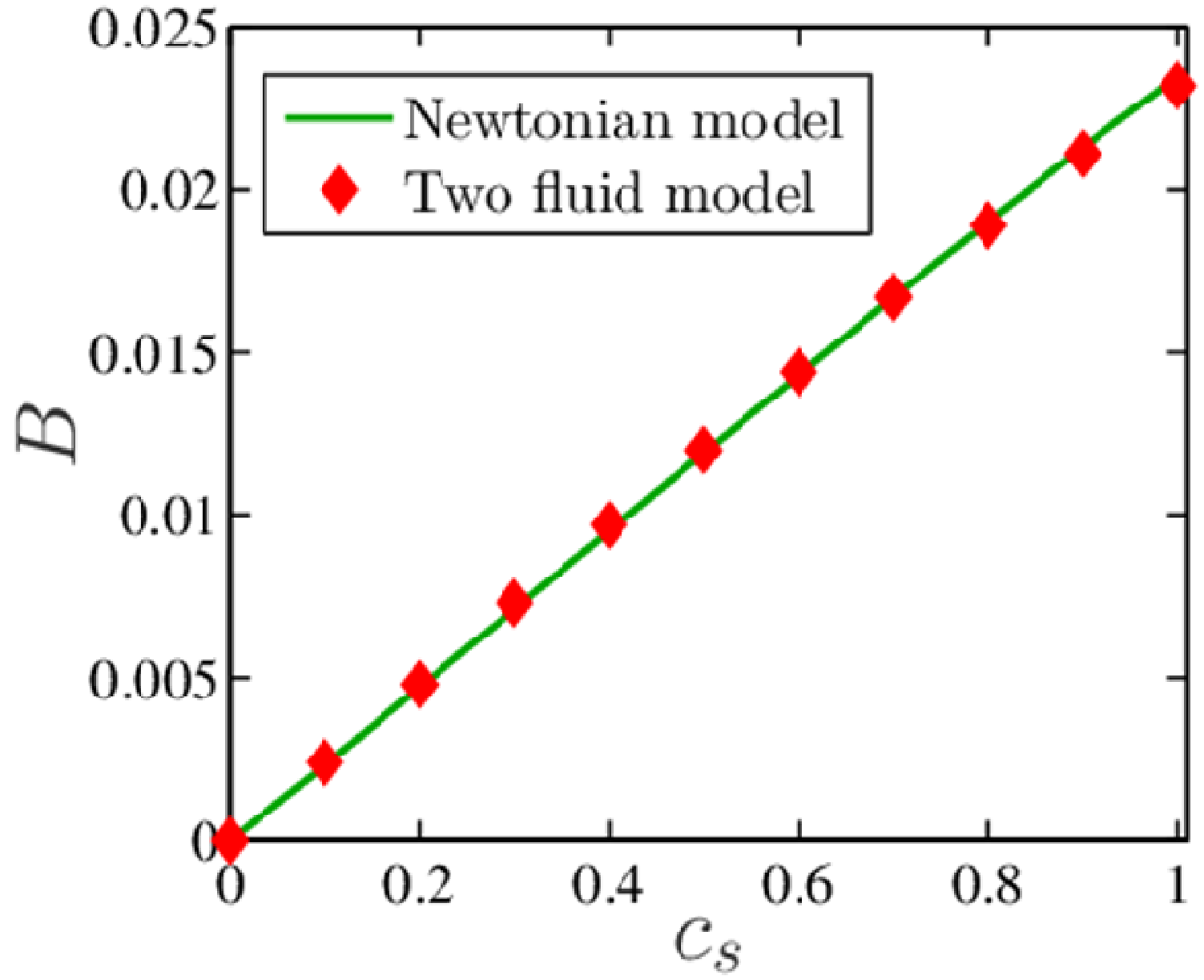

Figure 4: Validation of the present model with full-Newtonian model by assuming the values of $\mu_r = 1, De = 0$ and $h_1 = 0$.

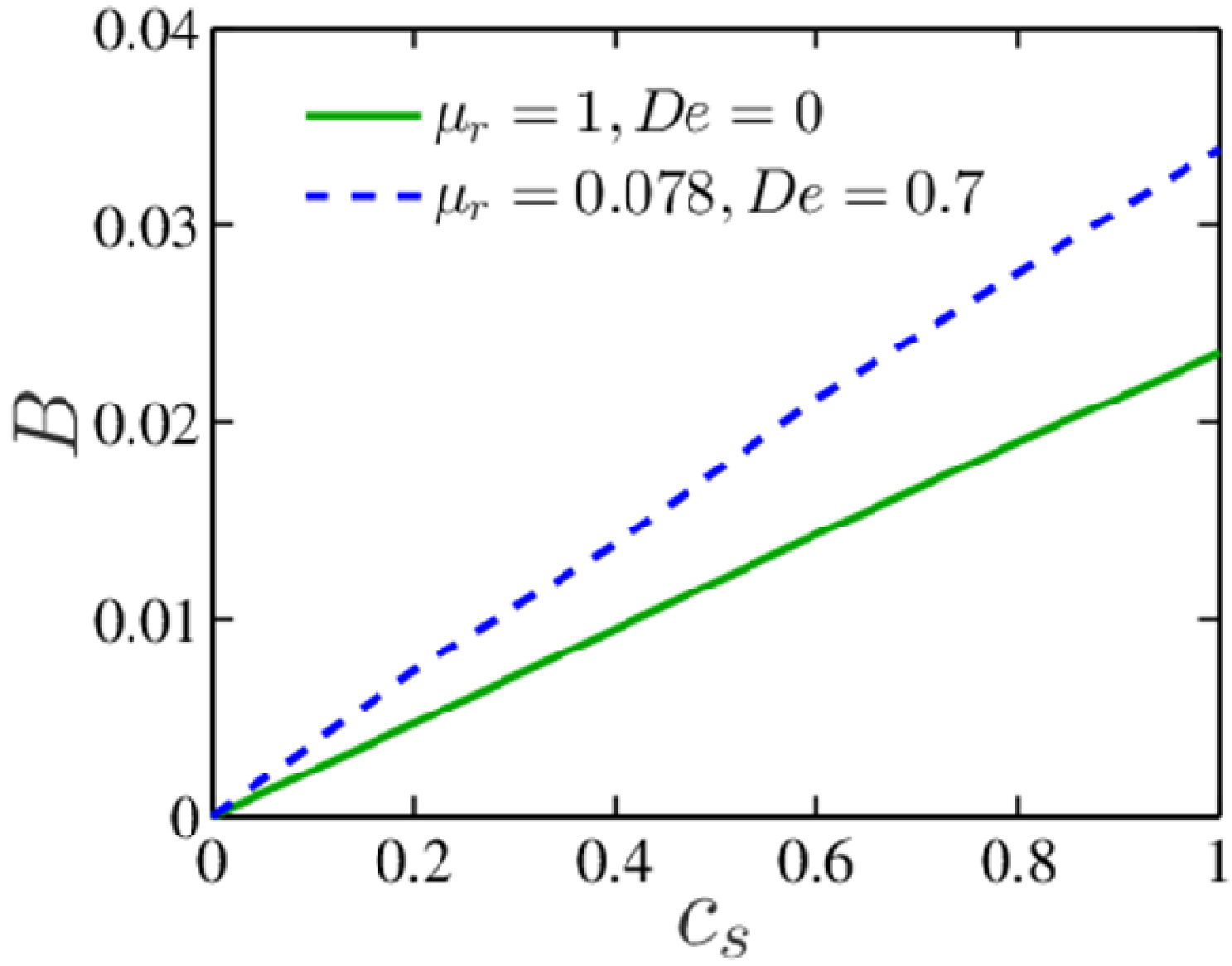

Figure 5: The difference in the value of streaming potential between single fluid $(\mu_r = 1, De = 0)$ and two-fluid model $(\mu_r = 0.078, De = 0.7)$.

After the validation and justification of the two-fluid model, we will now focus on the effects of physiological parameters on the induced streaming potential. Hematocrit fraction (Hct) denotes the percentage volume occupied by RBC in whole blood. The value of hematocrit fraction is generally found to be in the range of ~ 40% - 48% in healthy adult human beings (Divyashree & Gayathri 2018). Any significant change in the value of hematocrit fraction has been found to be directly linked to a person's health condition. Thus hematocrit fraction, thus, is an important physiological parameter. Hence, study of its effects and variation with health and diseased state has been an active area of research. Figure 6 shows the dependence of the induced streaming potential on the hematocrit fraction. The formation of cell free layer is dependent on the properties of RBC and other cells. In addition to that, the number of RBCs present in the whole blood has a direct influence on the thickness of cell free layer. Thus, it is important to take into account the cell free layer thickness when the effects of hematocrit fraction on streaming potential are analyzed. Variable $h_1$ takes into account the cell free layer thickness. For a given EGL thickness, higher the value of $h_1$ i.e. higher the distance of fluid-fluid interface from center line, thinner is the cell free layer.

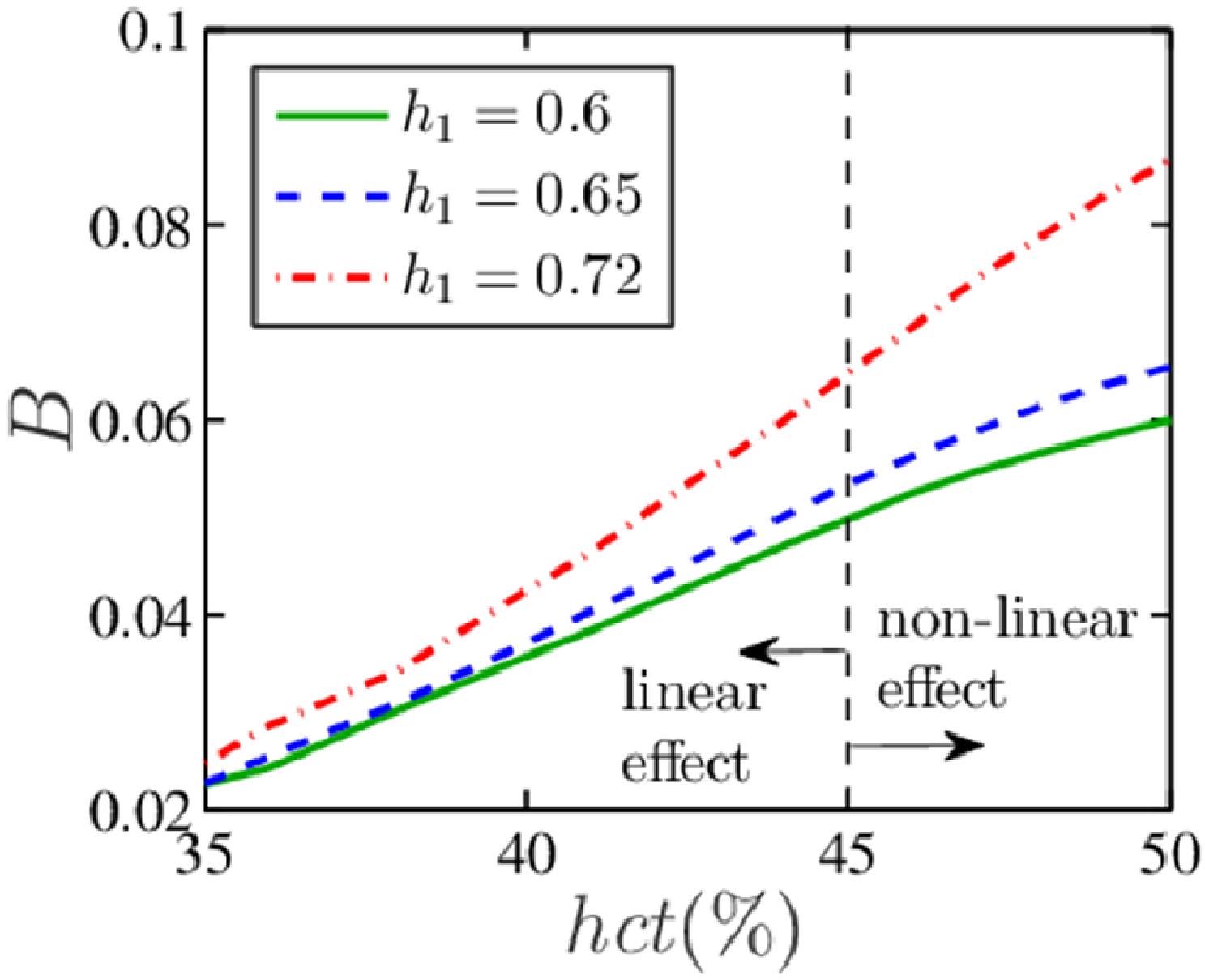

Figure 6: The dependence of streaming potential $(B)$ on hematocrit fraction $(hct)$.

The apparent viscosity of the whole blood is dependent on the hematocrit fraction as shown in the previous studies by (Chakraborty 2005; Das & Chakraborty 2006). From these reported studies, it may be inferred that the viscosity of whole blood, for different values of hematocrit fraction, range from 35% to 50%. The viscosity of blood plasma at 37 °C is found to be 1.34 mPa.s (Brust *et al.* 2013). This value is considered as reference in our study. Thus, effects of hematocrit fraction on streaming potential are accounted for by different values of the viscosity ratio.

The value of streaming potential is found to increase with increasing value of the hematocrit fraction with a linear rise, till hct of 45 %, beyond which it shows non-linear behavior with a decreasing slope as shown in figure 6. It is found that the streaming potential for hematocrit fraction of 50% is 2.5 to 3.5 times the streaming potential at 35% for different cell free layer thicknesses. It is shown in figure 7 (a) that the magnitude of the pressure gradient also increases with hematocrit fraction. This increase in magnitude of pressure gradient will enhance the advective motion of counter-ions present in the EDL which will induce more potential difference downstream. This is evident through the expression of electro-neutrality. The conduction current $\left\{ i_{cond.} = -B\left( \frac{c_+ + c_{l+}}{\gamma_+} + \frac{c_- + c_{l-}}{\gamma_-} \right) \right\}$ is dependent on the streaming potential and the ion concentration while the streaming current $\left\{ i_{stream} = u_3\left(c_+ - c_-\right) + u_2\left(c_{l+} - c_{l-}\right) \right\}$ is dependent on the ion concentration and the velocity. As the ion concentration is constant, the resulting streaming potential is directly dependent on the velocity which in turn is dependent on the pressure gradient. It is shown via figure 7 (b) that the increase in pressure gradient increases the streaming current because of the increase in advective motion of counter-ions and thus an increase in streaming potential is observed.

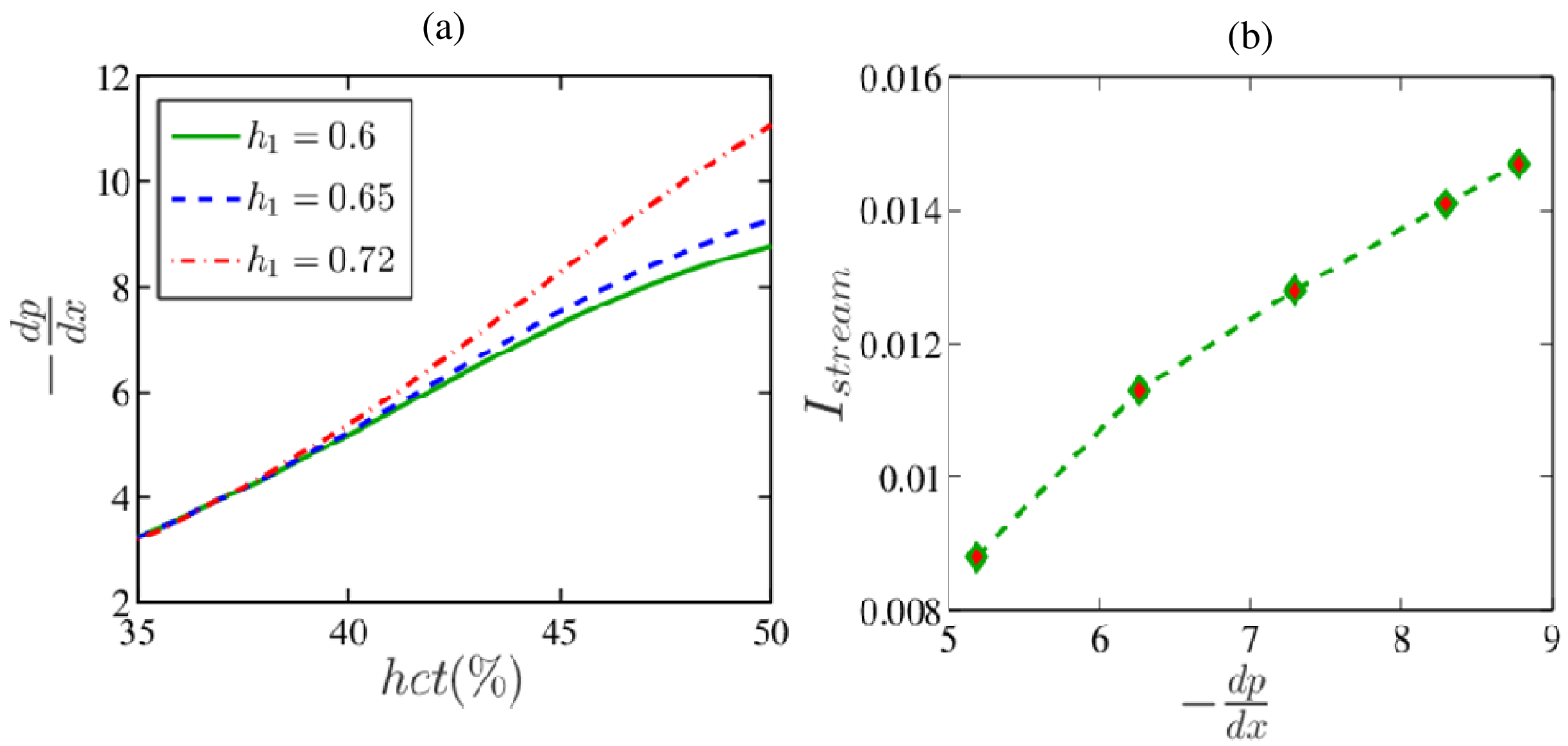

Figure 7: (a) The variation of the magnitude of pressure gradient with increasing hematocrit fraction $(hct)$, (b) The variation of streaming current with pressure gradient.

With the increase in hematocrit fraction, the whole blood becomes much more packed, resulting in an increased viscosity. This also means that the cell free layer thickness will reduce. Figure 8(a) shows the variation of streaming potential with $h_1$ (parameter taking the cell free layer thickness into account). It is observed that the value of streaming potential increases with $h_1$ i.e. decrease in cell free layer thickness. The pressure gradient also increases just as in case of increasing hematocrit fraction. The flow resistance is a direct function of cell free layer thickness, with lower resistance to the flow for thicker cell free layer and vice versa. This resistance is because of the increase in apparent viscosity of the blood. As the resistance increases, the pressure gradient should be increased to maintain same flow rate through the vessel (as shown in figure 8(b)) and this enhancement in the pressure gradient causes the streaming potential to increase as explained above. The streaming potential is shown to increase when $h_1$ is varied from $h_1 = 0$ (i.e. single fluid model) to $h_1 = 0.7$ (i.e. cell free layer thickness is 10% of vessel radius), with streaming potential for later case to be almost 3 times the streaming potential for single fluid model.

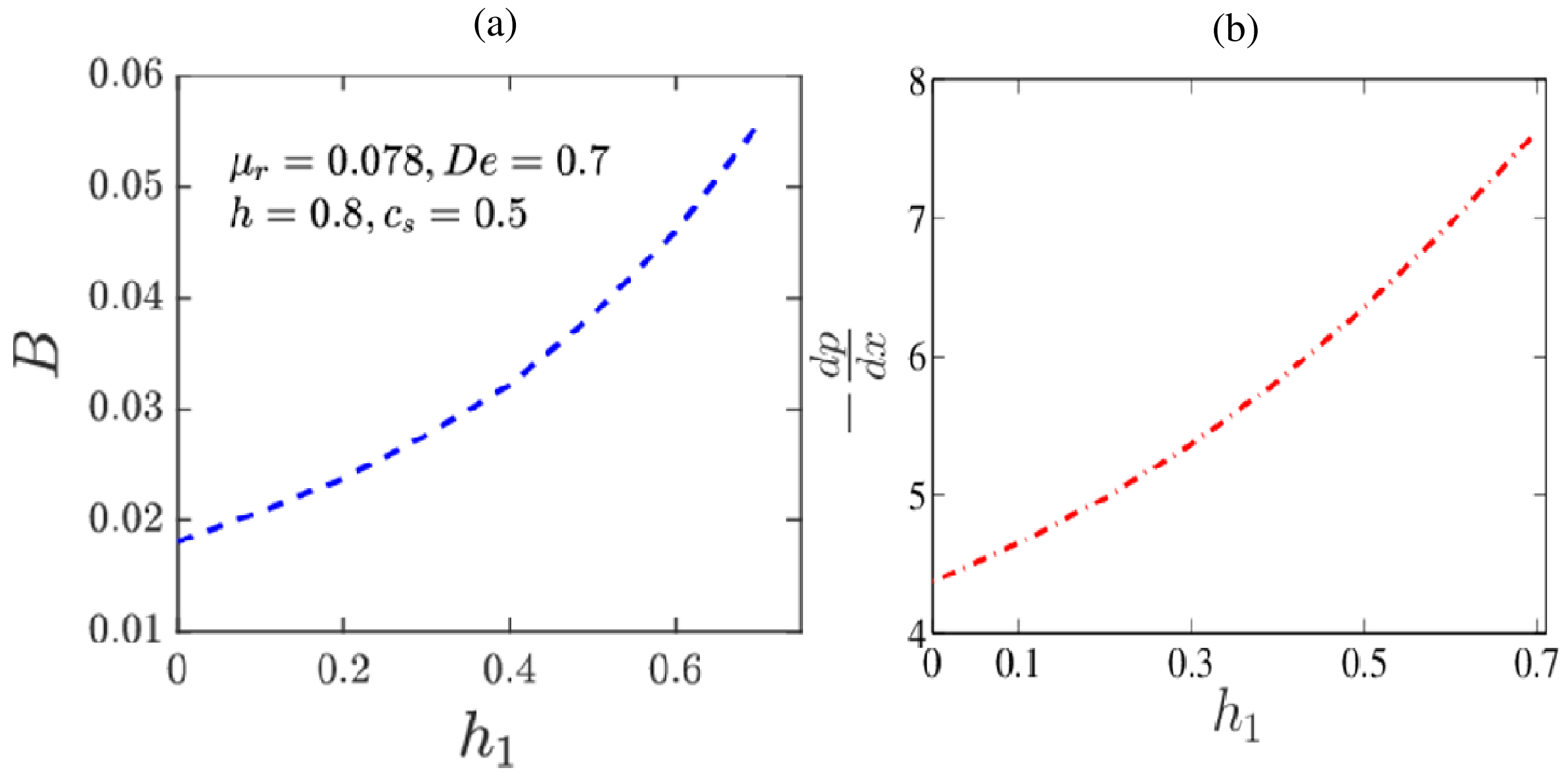

Figure 8: (a) The increase in the value of streaming potential with decreasing cell-free layer thickness (i.e. increasing $h_1$), (b) The variation of pressure gradient with decreasing cell-free layer thickness.

Just like hematocrit fraction, the thickness of EGL is also an important physiological parameter. Apart from EGL's function as mechano-transducer, EGL also plays an important role in many physiological phenomena like angiogenesis (Koumoutsakos *et al.* 2013). The EGL thickness is found to vary during angiogenesis and the sprouting usually takes place at locations where the EGL is absent, i.e. for zero EGL thickness. The mechanical, chemical and electrical signals resulting from the changes in EGL structure plays a vital role in the process of angiogenesis. Figure 9 (a) shows the variation of streaming potential with EGL thickness. Since the cell free layer thickness should vary because of variation of EGL thickness, the streaming potential for different values of EGL thickness are analyzed for $h_1 = 0.6$ (cell free layer thickness= 20% of vessel radius for $h = 0.8$) and $h_1 = 0.72$ (cell free layer thickness = 8% of vessel radius for $h = 0.8$), respectively. The results show that the streaming potential is more for thicker EGL and reduces as the EGL becomes thinner, with an almost zero potential when there is no EGL. This is because of our consideration of charge volume model where the fixed charge concentration is considered to be distributed throughout the poroelastic layer without any surface charge. Thus, with zero EGL thickness and absence of surface charge, the fixed charge concentration, $c_s$ becomes zero resulting in no electrochemical interaction with blood plasma. Thicker EGL implies that the solid volume fraction is more and thus the fixed charge concentration related to the solid matrix is more. This results in more electrochemical interaction. Along with this, the pressure gradient also increases with EGL thickness as shown in figure 9 (b). The combined effect of these two factors results in more augmentation in streaming potential.

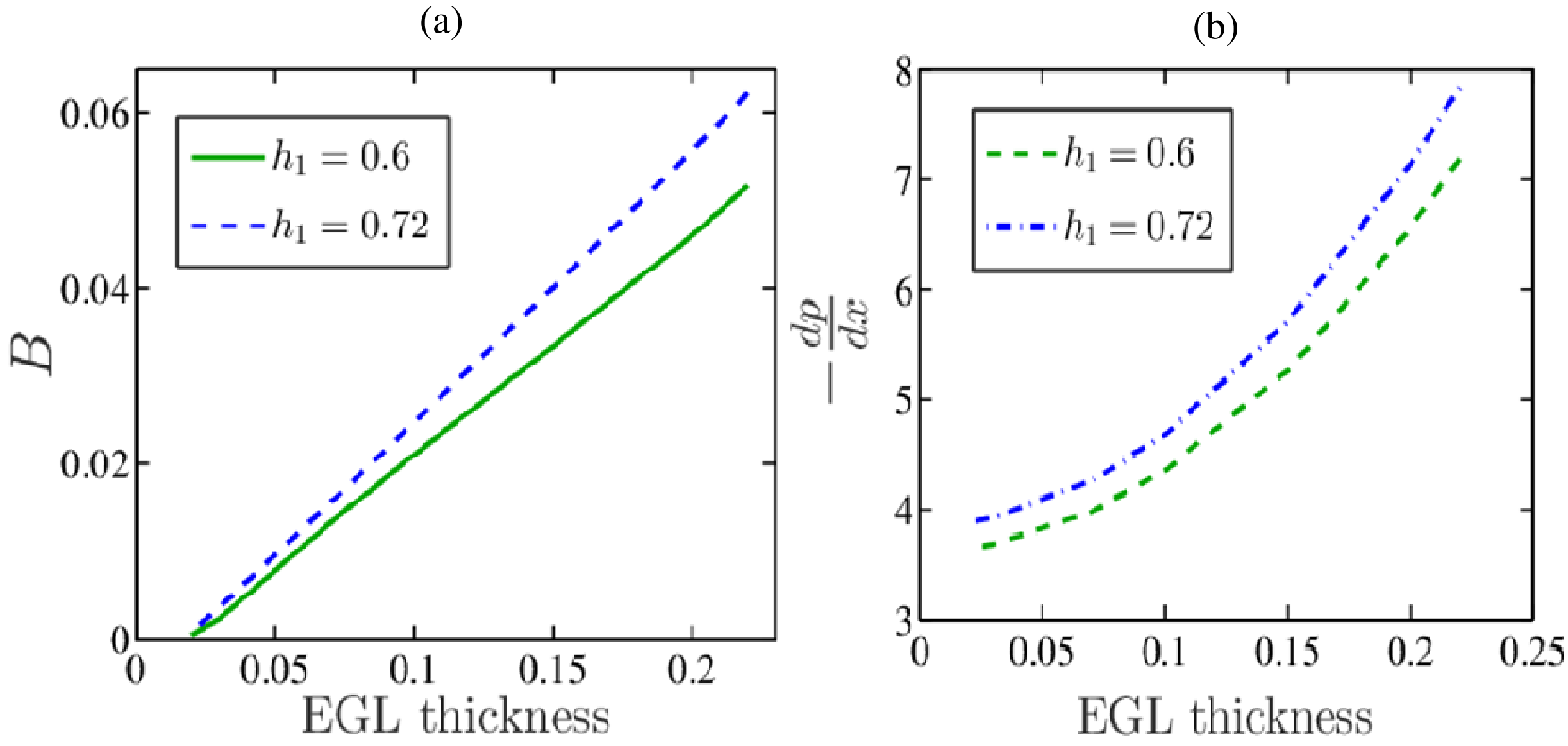

Figure 9: (a) The increase in the value of streaming potential with increasing EGL thickness, (b) The variation of pressure gradient with increasing EGL thickness.

Induced streaming potential in micro-fluidic systems has provided an option for powering the miniaturized devices. Utilizing the energy harnessed through this potential has been an active area of research. Powering the embedded bio-sensors and implantable medical devices (IMDs) has always been a challenging aspect. For micro-sensors and IMDs operating in a micro-vascular environment, batteries, although an established and reliable source of power (Bock *et al.* 2012; Nathan 2010; Scrosati 2011), are unsuitable due to their size, biocompatibility issues and potential toxicity (Amar *et al.* 2015). Quest for appropriate alternative sources of power within the physiological environment of human body is an active area of research. Typical powering devices such as bio-fuel cells (Bullen *et al.* 2006; Schmidt & Skarstad 2001; Wei & Liu 2008) and cells working on piezoelectricity (Gonzalez *et al.* 2002; Niu *et al.* n.d.), electrostatics (Miao *et al.* 2006; Tashiro *et al.* 2002), and electromagnetics (Goto *et al.* 1999) broadly work on environmental energy harvesting principle. Though the energy sources of these devices are sustainable and environment friendly, factors like lifetime (in case of bio fuels), limited implantable location (in case of piezoelectric devices), requirement of additional voltage source (in case of electrostatics based cells), electromagnetic radiation and highly complex design (in case of electromagnetic devices, present a limitation in their use. Along with these factors, size and biocompatibility issues add on to the limitations. Wireless charging of devices has also been proposed (Ma *et al.* 2018; Zhang *et al.* 2009), but the impact of electromagnetic signals on important physiological parameters and the range of charging prohibits their use in humans. Development of self powering devices which are safe, compact as well as sustainable is therefore of utmost requirement. The induced streaming potential because of blood flow over the endothelial glycocalyx layer provides an exciting alternative for powering these devices and sensors. Because this potential is induced due to flow of blood and not by any external source, harnessing it to power IMDs and bio sensors will be safe and also self powering. Along with this, it will also be suitable for powering devices implanted at complex microvascular environment. The value of streaming potential induced for blood flow with a hematocrit fraction of 45 % at 37 $^{o}$C is found to be in the tune of 0.2 mV/μm in our study. If the length is in millimeter scale the induced streaming potential will turn out to be of the order of Volt. Typical power requirement of the bio sensors and IMDs are in micro Watt to milli Watt range. Harnessing the induced streaming potential, which is of the order of 0.2 - 1 V, to generate power in the range of micro Watt to milli Watt, may turn out to be substantial for energizing the micro sensors and some IMDs. Even if the harnessed power turns out to be low, this can be used in sync with some other biocompatible power source which has low power output like bio-fuel cells. Thus, the induced streaming potential due to blood flow over the negatively charged EGL has its potential application in powering of embedded biosensors.

Along with its application as a power source, streaming potential and its dependence on various physiological parameters can be used for medical diagnostics and can also help in understanding some complex physiological phenomena. Determination of physiological parameters like hemoglobin level and hematocrit fraction is crucial in disease detection and is required prior to start some medical procedures. Conventional ways of determining these factors are complex as well as time consuming. To determine these parameters quickly and accurately, development of point of care devices is required (Maslow *et al.* 2016). These point-of-care devices can also be used for the purpose of medical diagnostics. Divyashree & Gayathri (2018), Sellahewa (2013) in their study related to dengue, highlighted the fact that the biggest challenge in combating high mortality rate in dengue is the lack of preliminary detection. They also observed that there is large difference in the hematocrit fraction in the diseased state as compared to the healthy state. In our study it is found out that there is a direct dependence of induced streaming potential on hematocrit fraction with streaming potential increasing with hematocrit fraction. This fluctuation is shown to be 2.5 to 3.5 times the value at normal hematocrit fraction. In diseased condition, especially in dengue, there is large variation in hematocrit fraction (the hematocrit fraction increases abruptly in a lethal dehydrated state) and capturing this variation by using the fluctuation in streaming potential in point of care device can help in preliminary detection of the disease in an easy and hassle free way.

Some physiological phenomena like angiogenesis are directly linked with the chemical and electrical signals associated with change in EGL structure. The pattern of angiogenesis, which is the process of formation of new blood vessels, is still a mystery to the researchers. Since EGL has a very important role in tumor vasculature, in extravasations and stress distribution (Borsig 2011; Kang *et al.* 2011; Maeda 2010; Pries *et al.* 2010), which heavily influences vascular endothelial growth factor (VEGF) signaling, study of blood flow over EGL is one of many important aspects required to unlock the mystery of angiogenesis pattern (Koumoutsakos *et al.* 2013). It has been found in our study that the streaming potential varies quite significantly with EGL thickness. The sprouting and cellular signaling is directly related with the EGL structure and thickness, with sprouting initiating from a spot where the EGL is absent. Thus the study of streaming potential with EGL thickness may prove to be one of the factors in understanding the angiogenesis pattern. Thus, along with the primary purpose of powering the IMDs, study of streaming potential and its correlation with physiological and EGL parameters have other potential applications.

## IV. CONCLUSIONS

We have brought out unique coupling of electromechanics and hydrodynamics over interfacial scales, so as to give rise to an induced streaming potential culminating out of blood flow in physiologically relevant microvessels. We incorporate the dependence of streaming potential on the physiological parameters by deploying a two-fluid model for pressure driven flow of blood where the blood plasma is assumed Newtonian and whole blood as non-Newtonian fluid. The EGL is modeled as poroelastic layer with fixed charge concentration associated with the solid matrix. The physiological parameters considered here have been found to significantly affect the induced streaming potential. By using these results, the streaming potential that is induced has been shown to have a possibility to be used as a safe and biocompatible power source for implantable medical devices and bio sensors. Along with this, its dependence and fluctuation with physiological parameters can play a significant role in medical diagnostics and in understanding of angiogenesis pattern which is a mystery for researchers working towards cancer cure. We envisage that the potential applications of the induced streaming potential in blood flows will be explored extensively in future.

## Acknowledgment

SC acknowledges Department of Science and Technology, Government of India, for Sir. J. C. Bose National Fellowship.

## APPENDIX

### A: Relationship between the tangential stress and normal stress

To establish the relationship between the tangential and normal stress, we first start with the equations related to the components of stress tensor

$$\left.\begin{aligned}
2\mu_1\frac{\partial u}{\partial x} &= f\left(\tau_{kk}\right)\tau_{xx}+\lambda_R\left(u\frac{\partial \tau_{xx}}{\partial x}+v\frac{\partial \tau_{xx}}{\partial y}-2\frac{\partial u}{\partial x}\tau_{xx}-2\frac{\partial u}{\partial y}\tau_{yx}\right)\\
2\mu_1\frac{\partial v}{\partial y} &= f\left(\tau_{kk}\right)\tau_{yy}+\lambda_R\left(u\frac{\partial \tau_{yy}}{\partial x}+v\frac{\partial \tau_{yy}}{\partial y}-2\frac{\partial v}{\partial x}\tau_{xy}-2\frac{\partial v}{\partial y}\tau_{yy}\right)\\
\mu_1\left(\frac{\partial u}{\partial y}+\frac{\partial v}{\partial x}\right) &= f\left(\tau_{kk}\right)\tau_{xy}+\lambda_R\left(u\frac{\partial \tau_{xy}}{\partial x}+v\frac{\partial \tau_{xy}}{\partial y}-\frac{\partial u}{\partial y}\tau_{yy}-\frac{\partial v}{\partial x}\tau_{xx}\right)
\end{aligned}\right\} \quad \text{(A1)}$$

For steady, incompressible, fully developed and unidirectional flow condition, three components of the stress tensors are simplified in the following way

$$\left.\begin{aligned} 0 &= f\left(\tau_{kk}\right)\tau_{xx} - 2\lambda_R \frac{\partial u}{\partial y}\tau_{yx} \\ 0 &= f\left(\tau_{kk}\right)\tau_{yy} \\ \mu_1 \frac{\partial u}{\partial y} &= f\left(\tau_{kk}\right)\tau_{xy} - \lambda_R \frac{\partial u}{\partial y}\tau_{yy} \end{aligned}\right\} \tag{A2}$$

Since $f\left(\tau_{kk}\right) = 1 + \frac{\in \lambda_R}{\mu_1}\tau_{kk}$ is non-zero, $\tau_{yy}$ has to be zero and $f\left(\tau_{kk}\right) = 1 + \frac{\in \lambda_R}{\mu_1}\tau_{xx}$. Now, we get the following two set of equations

$$f\left(\tau_{kk}\right)\tau_{xx} = 2\lambda_R \frac{\partial u}{\partial y}\tau_{yx} \tag{A3}$$

$$f\left(\tau_{kk}\right)\tau_{xy} = \mu_1 \frac{\partial u}{\partial y} \tag{A4}$$

Dividing equation (A3) by equation (A4), relationship $\tau_{xx}$ and $\tau_{xy}$ between is obtained

$$\tau_{xx} = 2\frac{\lambda_R}{\mu_1}\tau_{yx}^{\;2} \tag{A5}$$

Using this, the modified form of equation (A4) is written below

$$\frac{\partial u}{\partial y} = \frac{\tau_{xy}}{\mu_1} + \frac{2\in \lambda_R^{\;2}}{\mu_1^{\;3}}\tau_{yx}^{\;3} \tag{A6}$$

**B: Coefficients involved in the calculation of streaming potential**

After solving the equations using the four boundary conditions mentioned earlier, we get the value of $D_1$, $D_2$ and $B$ in terms of pressure gradient $A$ as follows

$$D_1 = AC_{11} + BC_{12}, D_2 = -AC_{13} - BC_{14},\; B = AC_{17} + A^3 C_{18} + C_{19} \tag{B1}$$

and $A$ is determined by solving the following cubic equation numerically

$$AC_{27} + A^3 C_{28} + C_{29} = 0 \tag{B2}$$

where the coefficients are as follows:

$$C_1 = \frac{\hat{\chi}_l}{\phi_f\left(\chi - 2\lambda\right)}, C_2 = \frac{c_s \hat{\chi}_l}{\phi_f \chi}, C_3 = \frac{1}{\chi}, C_4 = \frac{\hat{\chi}_l K}{\lambda}, C_5 = p \sinh\left(p h_1\right),$$

$$C_6 = \frac{1}{2} h_1^2 \left(\mu_r - 1\right), C_7 = \frac{1}{2} h_1^4 \in De^2 \mu_r^2, C_8 = \frac{\hat{\chi}_l K}{\lambda} \left\{\cosh\left(p h_1\right) - h_1 p \sinh\left(p h_1\right)\right\},$$

$$C_9 = 2\hat{\chi}_l \frac{\left\{K_1 e^{-p} + K_2 e^{p}\right\}}{e^m \phi_f\left(\chi - 2\lambda\right)} - \frac{c_s \hat{\chi}_l}{e^m \phi_f \chi}, C_{10} = -\frac{\hat{\chi}_l K p \sinh\left(p h\right)}{\lambda} + \frac{\hat{\chi}_l K p \sinh\left(p h_1\right)}{\lambda} - C_1 \left\{K_2 p e^{ph} - K_1 p e^{-ph}\right\},$$

$$C_{11} = -\frac{\left(h - C_{31}\sqrt{\chi} e^{\sqrt{\chi} h}\right)}{\left(\sqrt{\chi} C_{30} e^{\sqrt{\chi} h} + \sqrt{\chi} e^{-\sqrt{\chi} h}\right)}, C_{12} = \frac{C_{10} + C_9 \sqrt{\chi} e^{\sqrt{\chi} h}}{\left(\sqrt{\chi} C_{30} e^{\sqrt{\chi} h} + \sqrt{\chi} e^{-\sqrt{\chi} h}\right)},$$

$$C_{13} = C_{30} C_{11} - C_{31}, C_{14} = C_{30} C_{12} + C_9, C_{15} = \frac{1}{2} h^2 + \frac{1}{2} h_1^2 \left(\mu_r - 1\right) + \phi_f C_3,$$

$$C_{16} = -\frac{\hat{\chi}_l K \cosh(ph)}{\lambda} + \frac{\hat{\chi}_l Kp \sinh(ph_1) h}{\lambda} + \frac{\hat{\chi}_l K}{\lambda}\left\{\cosh(ph_1) - h_1 p \sinh(ph_1)\right\} - 2\hat{\chi}_l \left\{\frac{K_1 e^{-ph} + K_2 e^{ph}}{\chi - 2\lambda}\right\} + \frac{\hat{\chi}_l c_s}{\chi}$$

$$C_{17} = C_{15} + C_{13}\phi_f e^{mh} - C_{11}\left\{\frac{\phi_f e^{-mh}}{-C_{16} - \left(C_{14}\phi_f e^{mh}\right) + C_{12}\phi_f e^{-mh}}\right\}, C_{18} = \frac{C_7}{\left\{-C_{16} - \left(C_{14}\phi_f e^{mh}\right) + C_{12}\phi_f e^{-mh}\right\}},$$

$$C_{19} = \frac{1}{-C_{16} - \left(C_{14}\phi_f e^{mh}\right) + C_{12}\phi_f e^{-mh}},$$

$$C_{20} = \begin{bmatrix} 2K_2 m(m-p)\left\{\exp(C_{39}) - \exp(C_{42})\right\} + 2K_1 m(m+p)\left\{\exp(C_{40}) - \exp(C_{41})\right\} \\ +c_s\left(m^2 - p^2\right)\left\{\exp(C_{44}) - \exp(C_{43})\right\} \end{bmatrix} \frac{\exp(-C_{33})}{\left(m^2 - p^2\right)m},$$

$$C_{21} = \begin{bmatrix} 2K_2 m(m+p)\left\{\exp(C_{47}) - \exp(C_{52})\right\} + c_s\left(m^2 - p^2\right)\left\{\exp(C_{49}) - \exp(C_{48})\right\} \\ +2K_1 m(m-p)\left\{\exp(C_{50}) - \exp(C_{51})\right\} \end{bmatrix} \frac{\exp(-C_{33})}{\left(m^2 - p^2\right)m},$$

$$C_{22} = \frac{1}{24p}\begin{bmatrix} 48C_4C_5K\left(C_{55}h_1 - C_{57}h\right) + K\left(5K - 24C_4\right)\left(C_{55}C_{56} - C_{57}C_{58}\right) \\ +p\left(h_1 - h\right)\left(5K^2 + 20 - 24C_4K\right) + \frac{48}{p}C_4C_5K\left(C_{58} - C_{56}\right) + 4K\left(C_{55} - C_{57}\right)\left(12C_8 + 1\right) \end{bmatrix}$$

$$+\frac{exp(-C_{33})}{48p}\begin{bmatrix} (96C_1 + 10)K_1^{\ 2}\left\{exp(C_{54}) - exp(C_{53})\right\} + (96C_2 + 20c_s - 8 + 96C_1 c_s)K_1\left\{exp(C_{34}) - exp(C_{37})\right\} \\ +(96C_2 + 20c_s - 8 + 96C_1 c_s)K_2\left\{exp(C_{35}) - exp(C_{36})\right\} + (96C_1 + 10)K_2^{\ 2}\left\{exp(C_{45}) - exp(C_{38})\right\} \\ +(5c_s - 4 + 48C_2)\,exp(C_{33})c_s p(h-1) + exp(C_{33})\,p(h-1)\left\{40(1 + K_1K_2) + 384C_1K_1K_2\right\} \end{bmatrix},$$

$$C_{23} = \frac{2C_7K}{p}\left(C_{55} - C_{57}\right),$$

$$C_{24} = \frac{K}{p^3}\left[p^2\left(C_{55}h_1^2 - C_{57}h^2\right) + 2\left(C_{55} - C_{57}\right)\left(C_6p^2 + 1\right) - 2p\left(C_{56}h_1 - C_{58}h\right)\right]$$

$$+\left[C_3\exp(C_{33})c_s p(h-1) + 2C_3K_2\left\{\exp(C_{35}) - \exp(C_{36})\right\} + 2C_3K_1\left\{\exp(C_{34}) - \exp(C_{37})\right\}\right]\frac{\exp(-C_{33})}{p},$$

$$C_{25} = \frac{2K}{p}\left(C_{55} - C_{57}\right), C_{26} = C_{22} + C_{12}C_{21} - C_{14}C_{20}, C_{27} = C_{24} - C_{20}C_{13} + C_{21}C_{11} + C_{26}C_{17},$$

$$C_{28} = C_{26}C_{18} + C_{23}, C_{29} = C_{25} + C_{19}C_{26}, C_{30} = e^{-2m}, C_{31} = \frac{C_3}{e^m}, p = \sqrt{2\lambda}, m = \sqrt{\chi},$$

$$C_{32} = 2hm + 3hp + m + 2p,\ C_{33} = hm + 2hp + m + 2p, C_{34} = hm + hp + m + 2p, C_{35} = hm + 2hp + m + 3p,$$

$$C_{36} = hm + 3hp + m + 2p, C_{37} = hm + 2hp + m + p, C_{38} = hm + 2hp + m + 4p, C_{39} = 2hm + 3hp + m + 2p,$$

$$C_{40} = 2hm + hp + m + 2p, C_{41} = hm + 2hp + 2m + p, C_{42} = hm + 2hp + 2m + 3p, C_{43} = 2hm + 2hp + m + 2p,$$

$$C_{44} = hm + 2hp + 2m + 2p, C_{45} = hm + 4hp + m + 2p, C_{46} = C_{35}, C_{47} = hm + 2hp + 3p, C_{48} = hm + 2hp + 2p,$$

$$C_{49} = 2hp + m + 2p, C_{50} = hm + 2hp + p, C_{51} = hp + m + 2p, C_{52} = 3hp + m + 2p, C_{53} = hm + m + 2p,$$

$$C_{54} = hm + 2hp + m, C_{55} = sinh(ph_1), C_{56} = cosh(ph_1), C_{57} = sinh(ph), C_{58} = cosh(ph)$$